\documentclass[12pt,a4paper,twoside]{article}
\usepackage{epsfig}
\usepackage{baltlat1}
\usepackage{wrapfig}
\begin{document}
\ \ 
\vspace{0.5mm}
\setcounter{page}{1}
\vspace{8mm}
\titlehead{Baltic Astronomy, vol.13, -, 2004.}
\titleb{ Where is the 3$^{rd}$ subgroup of GRBs?}

\begin{authorl}
\authorb{I.~Horv\'ath}{1},
\authorb{A.~M\'esz\'aros}{2},
\authorb{L.~G.~Bal\'azs}{3} 
and
\authorb{Z.~Bagoly}{4}
\end{authorl}

\begin{addressl}
\addressb{1}{Dept. of Phys., 
Bolyai Military Univ. H1456 Budapest, POB 12, Hung}

\addressb{2}{Astr. Inst. Charles Univ. V Hole\v{s}ovi\v{c}k\'ach 2, 
CZ-180 00 Prague 8}

\addressb{3}{Konkoly Observatory, H-1525 Budapest, POB 67, Hungary}

\addressb{4}{Laboratory for Information Technology, E\"{o}tv\"{o}s
University, H-1117 Budapest, P\'azm\'any P. s.  1./A, Hungary}
\end{addressl}

\submitb{Received October 16, 2003}

\begin{abstract}
It is shown that in the duration-hardness plane the GRBs of the third
intermediate subgroup are well defined. Their durations are intermediate
(i.e. roughly between 2 and 10 seconds), but their hardnesses are 
the smallest. They are even softer than the long bursts.
\end{abstract}


\resthead{Where is the 3$^{rd}$ subgroup of GRBs?}{Horv\'ath et al.}

\sectionb{1}{Introduction}
 It is widely accepted that the short and long gamma-ray bursts (GRBs)
are really different phenomena (see, for example 
Norris et al. (2001) and the
references therein). 
In 1998 Horv\'ath made a trimodal fit of the BATSE
Catalog %(Meegan et al. 2001) 
bursts' duration distribution and found
a third subclass of GRBs' (Horv\'ath 1998).
Later papers (Mukherjee et al. 1998, Hakkila et al. 2000, 
Rajaniemi \& M\"ah\"onen 2002, Horv\'ath 2002) used
more parameters (e.g. peak fluxes, fluences,
hardness ratios), and in this multidimensional space studies
all of them confirmed that the third bursts population was
statistically necessary.

Bagoly et al. (1998) 
showed that in this high dimensional parameter spaces
only two main parameters were necessary to
characterize all the BATSE Catalog bursts' properties.
Hence, a two dimensional space looks like
a good characterisation of the GRB subgroups.
Therefore, here we use T$_{90}$ and the 
hardness $H32$ ratio in our newest analysis.
Figure 1. shows the observed BATSE bursts'
distribution on the $ \log (T_{90})-\log (H_{32})$ plane.

\vskip1mm
\begin{wrapfigure}{i}[0pt]{61mm}
\centerline{\psfig{figure=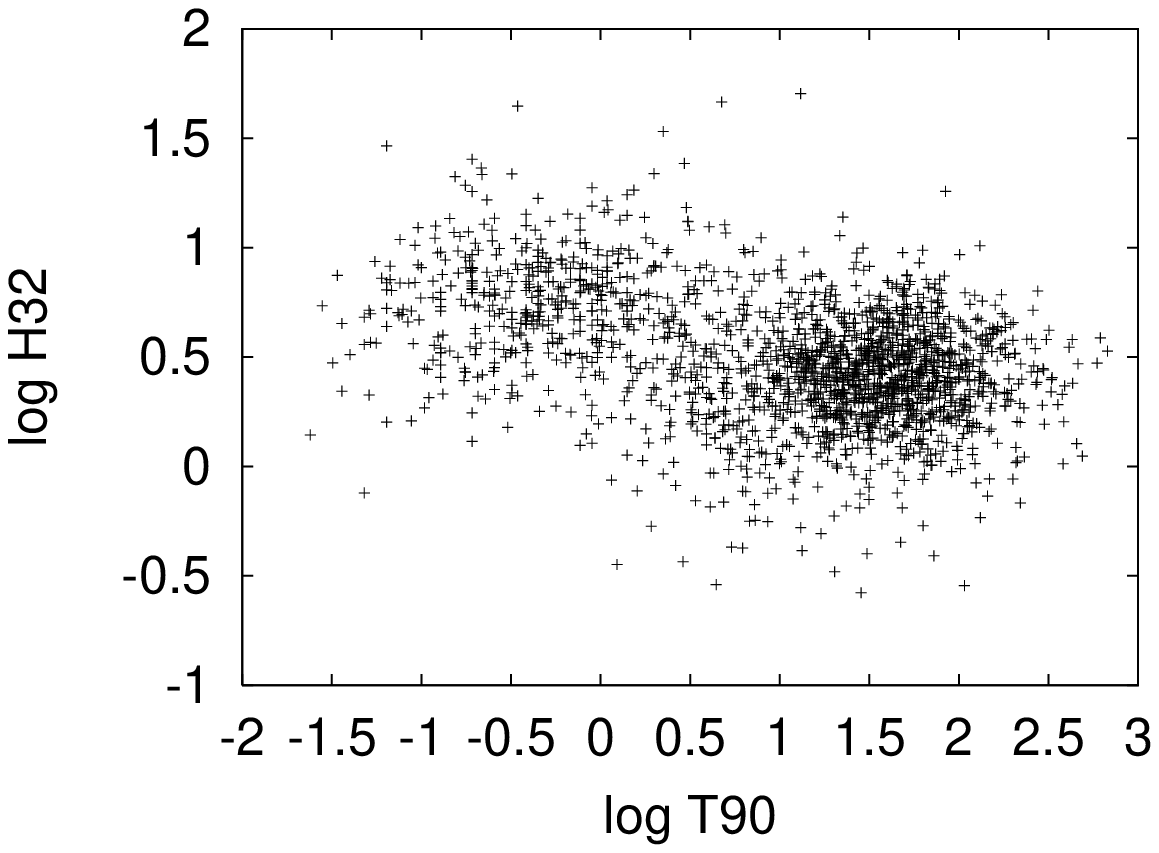,width=60truemm,angle=0,clip=}}
\captionb{1}{The values of $ \log T_{90}$ and $\log H32$ of
the 1929 BATSE GRBs.}
\end{wrapfigure}

\sectionb{2}{The fits}

This distribution can be fitted
at the first step by the sum of two two-dimensional 
normal distributions
in the $ \log T_{90}$ - $\log H32$ plane.
If one were fitting
{\it simultaneously} the values of $\log T_{90}$ and $\log H32$ 
by one single two-dimensional (bivariate) normal distribution, then 
the distribution would have five independent 
parameters (two means, two dispersions, and the 
correlation coefficient). The standard form of such
a bivariate distribution is given by
$$ f(x,y) dx dy =
\frac{N dx dy}{2 \pi \sigma_x \sigma_y \sqrt{1-r^2}}\times $$
\begin{equation}
\exp\left[-\frac{1}{2(1-r^2)}
\left(\frac{(x-a_x)^2}{\sigma_x^2} +
\frac{(y-a_y)^2}{\sigma_y^2} - \frac{C}
{\sigma_x \sigma_y}\right)\right]
 \;,
\end{equation}
where $x =\log T_{90}$, $y = \log H_{32}$, 
$C = 2r(x-a_x)(y-a_y)$, $a_x$, $a_y$ are the means,
$\sigma_x$, $\sigma_y$ are the dispersions, and $r$ is the correlation
coefficient. $N$ is the number of GRBs, and
$f(x,y) dx dy$ the theoretically expected numbers of GRBs at the
inf. surface at the $[x,y]$ plane given by intervals
$[x,(x+dx)]$ and $[y,(y+dy)]$. In other words, 
$(f(x,y)/N) dx dy$ defines the probability of finding a GRB at the
given inf. surface. 

%SMALL TABLE WRAPPED WITH TEXT
\begin{wrapfigure}{l}[0pt]{5.5cm}
\vbox{\footnotesize
\begin{tabular}{rrrrrrr}
\multicolumn{7}{c}{\parbox{5cm}{\baselineskip=8pt
~~~~{\smallbf Table 1.}{\small\ Best fit with two bivariate 
normal distributions for $ \log T_{90}$ and $\log H32$.  }}}\\
\tablerule
   $a_{x1}$ & -0.35      & & &   $a_{x2}$ & 1.47  &   \\
   $a_{y1}$ & 0.70      & & &   $a_{y2}$ & 0.39   & \\
   $\sigma_{x1}$ & 0.52  & & &  $\sigma_{x2}$ & 0.47 & \\
   $\sigma_{y1}$ & 0.33  & & &  $\sigma_{y2}$ & 0.24  &   \\
    $r_1$  & 0.1     & & &   $r_2$ & 0.1  &  \\
   W & 0.28             & & & &  &  \\

\tablerule
\end{tabular}
}
\end{wrapfigure}

This fit was done by a standard search for 
11 parameters with $N=1929$ 
measured points. 
Each GRB defines a point in the $x, y$ plane
with coordinates $x_i, y_i$ $(i = 1,2,...,N)$. The theoretical curve
$f_2(x,y, a_{xk}, a_{yk},\sigma_{xk}, \sigma_{yk}, r_k, W)$
($k=1,2$) is a sum of two normal distributions as given in
Eq.1. The normalization constant of the first [second] term is $NW$ [$N(1-W)$],
where $W$ is the weight of the first normal distribution. 
For the first (second) term
the parameters are $a_{x1}, a_{y1},\sigma_{x1}, \sigma_{y1}, r_1$
($a_{x2}, a_{y2},\sigma_{x2}, \sigma_{y2}, r_2$).

We obtain the best fit to the 11 parameters through a maximum likelihood
(ML) estimation.
We search for the maximum of the formula
%Equ.5
\begin{equation}
L_2 = \sum_{i=1}^{N} \ln f_2(x_i, y_i)
\end{equation}
using a simplex numerical procedure; the index "2" in $L_2$ shows 
that we have a sum of two log-normal distributions of type given by Eq.1.

The results of this fit are shown in Table 1. 
One can calculate a density distribution of the observed
data on the log(T$_{90}$)-log(H$_{32}$) plane.
The values of Table 1 define the theoretical distribution of GRBs,
if there is any theory, which suggests lognormal H32 and $T_{90}$
distributions,
in the $ T_{90}$ - $ H32$ plane under the assumption that there
are {\it only} two subgroups. %Nevertheless, we already know 
%many papers have suggested that there are three subgroups (or may be more).  

\vskip1mm
%SINGLE CENTERED WRAPPED TEXT FIGURE
\begin{wrapfigure}{i}[0pt]{61mm}
\centerline{\psfig{figure=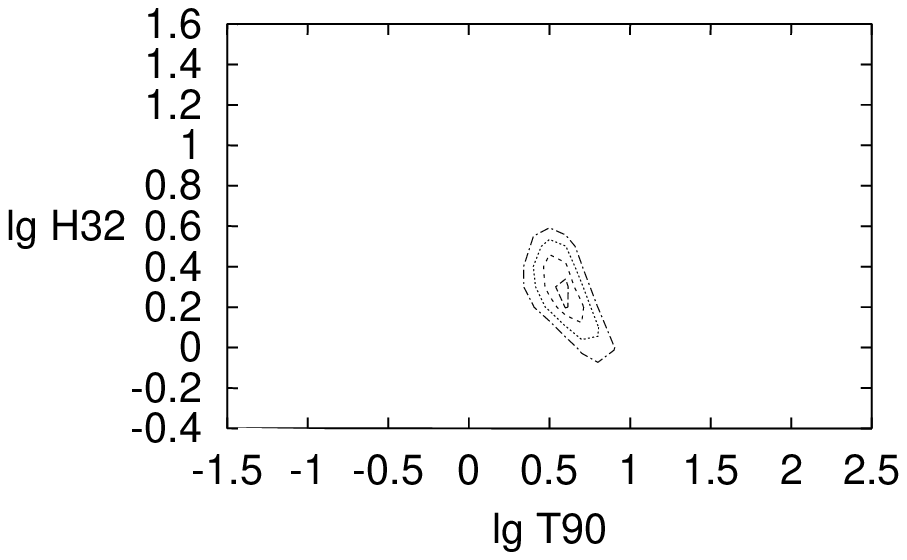,width=80truemm,angle=0,clip=}}
\captionb{1}{Departure of the numbers of GRBs
from the theoretical values given by the best fit collected at Table 
1.}
\end{wrapfigure}

Hence, one can make the difference between the actual distribution
and the theoretical distribution. 
If there were only two components, which distributed lognormally,
one would find only some noise being
%And of course, the noise is 
proportional to the square root of
the actual value of the burst density function.
Surprisingly the biggest deviation is not there where the
observed (or the theoretical) burst density function is
the biggest, not even the dense core of the two subgroup
population, rather then between the two groups in $T_{90}$
but in H32 softer then either groups.

This deviation is shown by Figure 2.
The widest contour has the half intensity then the
highest deviation.
In order to be sure that this difference
is not a chance and is not given by the random noise, %around the theoretical
 we proceeded as follows. 
We take this part of the $\log T_{90}$ - $\log H32$ plane. 
This part contains 145 bursts. The integral over this area
of the theoretical distribution gives 63.

Depends on which number is the Poisson parameter
one can calculate the probability $10^{-17}$-$10^{-14}$. 
Of course there are so many different area and shapes, which we
can probe. However if we take 10 different shapes 100 different sizes
and try 1000 different positions in the plane the probability is
still very low. Therefore one can say the deviation which the
Figure 2. shows is very unlikely can cause by chance.

Unfortunately, the possible third group members are mixed
with the long ones (partially also with the short ones, too).

\sectionb{3}{Conclusions}

We have argued that both the duration and also the hardness
should be distributed log-normally - of course, in any subclass
separately. We also provided in the $\log T_{90}$ - $\log H32$
plane a fit with the sum of two bivariate normal distributions.
The idea for this fitting was given by the observational fact that
the short and long subclasses are different both at hardnesses and
durations. Finally we obtained the difference between the
actual distribution of GRBs and the theoretical one. The difference, not
being a Poissonian noise, shows another subgroup, 
the third subgroup.

As the result we obtained the intermediate subclass having approximately
the same number of bursts than it was previously predicted
( Horv\'ath 1998, Mukherjee et al. 1998, Hakkila et al. 2000, 
Rajaniemi \& M\"ah\"onen 2002 and Horv\'ath 2002). 
We also confirm that
 their hardnesses are  low.

ACKNOWLEDGMENTS.\ This research was 
supported by Czech Research Grant J13/98: 113200004 (A.M.),
and by OTKA  T034549.
\goodbreak

\References

Bagoly, Z., M\'esz\'aros, A.,
Horv\'ath, I., Bal\'azs, L.G. \& M\'esz\'aros, P. 1998, ApJ, 498, 342

Bal\'azs, L.G., M\'esz\'aros, A., Horv\'ath, I. \& Vavrek, R. 2003, 
A\&A, 138, 417

Hakkila, J., et al. 2000, ApJ, 538, 165

Horv\'ath, I. 1998, ApJ, 508, 757

Horv\'ath, I. 2002, A\&A, 392, 791

Mukherjee, S., et al. 1998, ApJ, 508, 314

Norris, J.P., et al. 
 2001, in GRBs in the Afterglow Era, Proc.
 p. 40

Rajaniemi, H.J., \& M\"ah\"onen, P. 2002, ApJ, 566, 
202

\end{document}